\documentclass[aps,pre,superscriptaddress,showpacs,preprintnumbers]{revtex4-2}
\usepackage[dvipsnames]{xcolor}
\usepackage{soul}
\usepackage{graphicx}
\usepackage{dcolumn}
\usepackage{bm}
\usepackage[utf8]{inputenc}
\usepackage[hypertexnames=false,pdftex,colorlinks=true,urlcolor=blue,
            linkcolor=Blue,citecolor=RedViolet]{hyperref} 
\usepackage{braket}
\usepackage{physics}
\usepackage{mathtools}
\usepackage{tikz-cd}
\usepackage{dsfont}
\usepackage{scalerel}
\usepackage{ulem}
\usepackage{cancel}
\usepackage{mathbbol}

\newcommand{\beq}{\begin{equation}}
\newcommand{\eeq}{\end{equation}}
\newcommand{\bse}{\begin{subequations}}
\newcommand{\ese}{\end{subequations}}
\newcommand{\bea}{\begin{eqnarray}}
\newcommand{\eea}{\end{eqnarray}}

\def\ie{\textit{i.e., }}

\def\lala{\langle\langle}
\def\rara{\rangle\rangle}
\def\calN{{\cal N}}
\def\calM{{\cal M}}

\def\AA{\mathbb{A}}

\def\00{\mathbb{0}}

\def\I2n{\mathbb{l}}
\def\I2n{\mathbb{I}}
\def\JJ{\mathbb{J}}

\def\AA{\mathbb{A}}

\def\hE{\hat{E}}

\def\hG{\hat G}

\def\lp{l^\prime}

\def\matPhi{\hat{\mathbb{\Phi}}}
\def\hrho{\hat{\rho}}

\def\hK{\hat{K}}
\def\hH{\hat{H}}
\def\hV{\hat{V}}

\def\hU{\hat{U}}

\def\hid{\hat{\mathbb{1}}}

\def\calL{{\cal L}}
\def\calN{{\cal N}}
\def\calH{{\cal H}}
\def\calD{{\cal D}}

\def\hM{\hat M}
\def\hKb{\hat{\bar{K}}}
\begin{document}
\title{Quantum thermodynamics for general bipartite interacting autonomous systems}
\author{Fabricio Toscano}  
\email{toscano@if.ufrj.br}
\affiliation{Instituto de F\'isica, Universidade Federal do Rio de Janeiro, 21941-972, Rio de Janeiro, Brazil}
\author{Diego A. Wisniacki}
\email{wisniacki@df.uba.ar}
\affiliation{Departamento de F\'isica, FCEyN, UBA, Pabell\'on 1 Ciudad Universit\'aria, C1428EGA Buenos Aires, Argentina}
\date{\today}
\begin{abstract}
 The internal energy of individual subsystems is not well defined in interacting quantum systems, leading to ambiguities in the definition of thermodynamic quantities. 
 Applying the Schmidt basis formalism to general bipartite autonomous quantum systems, we demonstrate that the master equation describing subsystem evolution adheres to the principle of minimal dissipation. This enables to define internal energy of each subsystem in a consistent way.  Moreover, by utilizing general aspects of open quantum systems, we show that this master equation is unique. We analyze heat and work for each subsystem as derived from this formalism, providing deeper insights into the thermodynamics of interacting quantum systems.
\end{abstract}
\maketitle

\section{Introduction} 
\par

One of the most intriguing endeavors in modern physics is to firmly establish the emergence of thermodynamics from quantum mechanics as a foundational theory \cite{Gemmer2009}. 
This involves defining thermodynamic quantities within the quantum mechanical framework in a way that their behavior mirrors that of their classical counterparts, accurately describing scenarios where classical mechanics applies. However, the true significance of this program lies in gaining a thermodynamic understanding of genuinely quantum systems, particularly small-scale systems, as thermodynamic quantities derived from quantum formalism must also apply to them \cite{alicki2018introduction,deffner}. This field of quantum thermodynamics has gained momentum due to major advancements in quantum devices for information processing and many-body experiments \cite{islam2015measuring,eisert2015quantum}.

\par
Among the important quantities to be defined in quantum thermodynamics are heat and work. These define, for example, the first law of thermodynamics as part of quantum formalism. Several attempts have been made in the recent past, but without a conclusive result, particularly when considering two interactive systems \cite{weimer2008local,alipour2016correlations,strasberg2017quantum,rivas2020strong}. 
The root problem is how to define the internal energy of each subsystem consistently in a way that the sum of each internal energy is the total energy of the whole system, \i.e. the energy additivity for the bipartition. Consistently means that the internal energies of the subsystems must assume the influence of the interaction quantum Hamiltonian term, as well as the free evolution Hamiltonians in each part of the system.
\par
For arbitrary  bipartite autonomous quantum systems, Malavazi and Brito~\cite{malavazi2022schmidt} propose an approach to this problem based on the Schmidt decomposition for the evolution of a product of initial pure states
\cite{footnote0}. Indeed, they show that this formalism allows for the derivation of effective local operators that characterize the internal energies of each constituent system. The local Hamiltonians of their proposal naturally satisfy the notion of energy additivity for the bipartition, include the local free Hamiltonians of each part and also depend on the interaction 
quantum Hamiltonian term. Since the local effective Hamiltonians automatically become time dependent, there is a clear rule for defining work and hence heat in this proposal, so that the first law of thermodynamics is established \cite{alicki2018introduction}. The proposal has the advantage of being exact not relying on any restrictive approximation. 
However, it is not clear if there is another alternative way to define local effective Hamiltonians that verifies all the required properties. Furthermore, although these local effective Hamiltonians appear in the description of the dynamics of the reduced states, it is not clear how these operators appear
from an open quantum systems theory perspective, where the description of the dynamics of the reduced state of one of the subsystems consists in considering the other as an environment.
\par
In this paper, we show that the appropriate theoretical framework for addressing these issues is to consider the recent results in open quantum systems and quantum thermodynamics given in \cite{HaydenSorce2022} and \cite{colla2022open} respectively. Indeed, in \cite{HaydenSorce2022} was shown that, for finite dimensional quantum bipartite systems whose evolution is described by a completely 
positive dynamical map, the reduced dynamics is generically described by a
master equation in the Lindblad form, with time-dependent Linblad's operators. Furthermore, they showed that within all the equivalent unitary parts of the master equation, provided by the group of symmetries that leave the whole master equation invariant, there is a unique one, associated with a special local effective Hamiltonian, called canonical, that comes together with  a non-unitary part of the master equation, namely the dissipative part, of minimum size of a suitable norm.
Thus, in Ref. \cite{colla2022open} the principle of minimum dissipation was developed which identifies the canonical Hamiltonian of a master equation as the appropriate Hermitian operator to consistently define the local energy. In this way, work and heat in the open quantum system can be defined unambiguously~\cite{colla2022open,alicki2018introduction}.
\par 
In this work, we use the Lindblad form of the master equation 
for each subsystem that comes from the Schmidt decomposition for the evolution, to show that the local effective Hamiltonians of Malavazi and Brito ~\cite{malavazi2022schmidt} are indeed the canonical Hamiltonians, thus 
the dissipation part of the master equation obeys the principle of minimal dissipation. Furthermore, we demonstrate, using the general theory of open quantum systems, that the master equation from the Schmidt decomposition of the evolution, together with its group of symmetries, is the unique possible exact master equation 
that describe the evolution of the reduced states. Finally, we study the heat and work associated with each of the parts as derived from this formalism.
\par
The paper is organized as follows. In Sec. \ref{sec:intersys}, we present the considered system: a general interacting bipartite system and the evolution
of its parts given by a exact Lindblad master equation derived from the Schmidt decomposition.
In Sec. \ref{sec:ham}, 
we demonstrate that the dissipative part of this evolution is minimized, leading to the unitary component being determined by the so-called canonical Hamiltonian. In Sec. \ref{sec:open}, we further establish, within the framework of open quantum systems, that the exact master equation from the Schmidt decomposition is the unique description for the evolution. Building on this uniqueness in Sec. \ref{sec:thermo}, we explore the quantum thermodynamics of the system, specifically detailing the definitions and behavior of heat and work.
We present our conclusions in Sec. \ref{sec:conclusions}
%
\section{The interacting systems considered} 
\label{sec:intersys}
\par
Let us consider two interacting systems described by an autonomous Hamiltonian 
of the general form,
\beq
\label{eq:defH}
\hH=\hH_1\otimes \hid_2+\hid_1\otimes \hH_2+\lambda \hV,
\eeq 
with $\hH_j$, $j=1,2$, the free Hamiltonians, $\hV$ an interaction term and $\lambda$ a real number.
The evolution of an arbitrary separable pure state $\ket{\phi^{(1)}_0}\ket{\phi^{(2)}_0}$ is,
\bea
\label{eq:SchDecOrg}
\ket{\Psi(t)}&=&e^{-\imath \hH t }\ket{\phi^{(1)}_0}\ket{\phi^{(2)}_0} \nonumber\\
&=&\sum_{k=0}^{M-1} s_k(t) \ket{\phi^{(1)}_k(t)}\ket{\phi^{(2)}_k(t)},
\eea
where we consider $\hbar=1$. The last equality is the Schmidt decomposition of the evolved pure state
with  $\{\ket{\phi^{(j)}_l(t)}\}_{l=0,\ldots,N_j-1}$, 
$j=1,2$, the orthonormal bases of the Hilbert spaces ${\cal H}_j$, of dimension $N_j=\dim({\cal H}_j)$,
that are eigenstates of reduced states $\hrho^{(j)}(t)=\Tr_j(\dyad{\Psi(t)})$ with real eigenvalues $s_k(t)> 0$
for $l=0\ldots M-1$, $M\leq\min(N_1,N_2)$, and null eigenvalues for $l>M-1$. Note that the number $M$ it depends on time, in particular for $t=0$, $M=1$, so we have only one Schmidt coefficient $s_0(0)=1$.
\par
Alternatively, we can rewrite the Schmidt decomposition in~\eqref{eq:SchDecOrg}
as,
\bea
\ket{\Psi(t)}&=&e^{-\imath \hH t }\ket{\phi^{(1)}_0}\ket{\phi^{(2)}_0}\nonumber\\
&=&\hU^{(1)}(t)\otimes \hU^{(2)}(t)\sum_{l=0}^{M-1} s_l(t) \ket{\phi^{(1)}_l}\ket{\phi^{(2)}_l},
\label{eq:SchDecBritosimpl}
\eea
where we define $ \ket{\phi^{(j)}_l(0)}= \ket{\phi^{(j)}_l}$ for $l=0,\ldots,N_j-1$,
and $\hU^{(j)}(t)$ are the local unitary evolutions that transform the bases $\{\ket{\phi_l^{(j)}}\}_{l=0,\ldots,N_j-1}$ 
into $\{\ket{\phi^{(j)}_l(t)}\}_{l=0,\ldots,N_j-1}$, whose matrix elements are
$\matrixel{\phi_l^{(j)}}{\hU^{(j)}(t)}{\phi_{\lp}^{(j)}}=\braket{\phi_l^{(j)}}{\phi_{\lp}^{(j)}(t)}$.
The generators of these unitary operations, $\imath \dv{\hU^{(j)}(t)}{t}=\hH^{(j)}(t)\hU^{(j)}(t)$, are the Hamiltonians,
\beq
\hH^{(j)}(t)=\sum_{l=0}^{N_j-1}\imath \dv{\ket{\phi^{(j)}_l(t)}}{t}\bra{\phi^{(j)}_l(t)}.
\label{eq:defHjt}
\eeq
\par
Defining the transition operators 
$\hG^{(j)}_{l\lp}(t)\equiv\dyad{\phi^{(j)}_l(t)}{\phi^{(j)}_{\lp}(t)}$,
we have that $\hG^{(j)}_{ll}(t)=\hG^{(j)}_{l\lp}(t)\hG^{(j)\dagger}_{l\lp}(t)$, so the reduced states results, 
\bse
\bea
\hrho^{(j)}(t)&=&\sum_{l=0}^{M-1}s^2_l(t)\hG^{(j)}_{ll}(t)
\label{eq:hrhojta}\\
&=&\sum_{l=0}^{M-1}\hG^{(j)}_{l\lp}(t)s^2_l(t)\hG^{(j)\dagger}_{l\lp}(t).
\label{eq:hrhojt}
\eea
\ese
Taking the time derivative of the expression in~\eqref{eq:hrhojt} and using Eq.~\eqref{eq:defHjt} we arrive 
to the master equation,
\bea
\dv{\hrho_j(t)}{t}
&=&-\imath[\hH^{(j)}(t),\hrho_j(t)]+
 \sum_{l=0}^{M-1}\dv{s^2_l(t)}{t} \hG^{(j)}_{ll}(t).
  \label{eq:maserteqrho}
\eea
 The non-unitary term in the master equation Eq.~\eqref{eq:maserteqrho} can be written in the Lindblad's form (see \cite{malavazi2022energetic}); i.e.
\begin{widetext}
\bse
\label{eq:derirhoj}
\bea
\sum_{k=0}^{M-1}\dv{s^2_k(t)}{t} \hG^{(j)}_{ll}(t)&=&\sum_{l=0}^{M-1}\sum_{\lp=0}^{M-1}g_{l\lp}(t)\;\hG^{(j)}_{l\lp}(t)\hrho^{(j)}(t)\hG^{(j)\dagger}_{l\lp}(t)\\
&=&\sum_{l=0}^{M-1}\sum_{\lp=0}^{M-1}g_{l\lp}(t)\;\left(\hG^{(j)}_{l\lp}(t)\hrho^{(j)}(t)\hG^{(j)\dagger}_{l\lp}(t)
-\frac{1}{2}\{\hG^{(j)\dagger}_{l\lp}(t)\hG^{(j)}_{l\lp}(t),\hrho^{(j)}(t)\}\right),
\label{eq:derirhojb}
\eea 
\ese
\end{widetext}
where we define,
\beq
\label{eq:defgllpt}
g_{l\lp}(t)\equiv\frac{1}{Ms_{\lp}^2(t)}\dv{s^2_{l}(t)}{t},
\eeq
and in~\eqref{eq:derirhojb} we used that, 
\bea
\sum_{l=0}^{M-1}\sum_{\lp=0}^{M-1}g_{l\lp}(t)\frac{1}{2}\{\hG^{(j)\dagger}_{l\lp}(t)\hG^{(j)}_{l\lp}(t),\hrho_j(t)\}&=&0.
\label{eq:secondtermzero}
\eea
\par 
This last equation is obtained using,
\bea
\sum_{l=0}^{M-1}\sum_{\lp=0}^{M-1}g_{l\lp}(t)\hG^{(j)\dagger}_{l\lp}(t)\hG^{(j)}_{l\lp}(t)&=&0,
\label{eq:secondtermzero2}
\eea
and the normalization condition on the reduced states, \ie $\Tr(\hrho_j(t))=1$, so $\sum_{k=0}^{M-1}s_k^2(t)=1$ $\forall t$, then  the Schmidt coefficients satisfy,
\bea
\sum_{k=0}^{M-1}\dv{s^2_k(t)}{t}=0.
\eea
When $t=0$, $s_k(0)=\delta_{0k}$ so $\left.\dv{s_0(t)}{t}\right|_{t=0}=0$.
\par
 
As in any master equation of the Lindblad form, the r.h.s. of Eq.~\eqref{eq:maserteqrho}
is invariant if we apply the  inhomogeneous transformations,
 \begin{widetext}
\bse
\label{eq:transfLinbop}
\bea
\label{eq:transfLinbopa}
\hG^{(j)}_{l\lp}(t)&\rightarrow&\hE^{(j)}_{l\lp}(t)= \hG^{(j)}_{l\lp}(t)+\alpha^{(j)}_{l\lp}(t)\hid_j,\\
\hH^{(j)}(t)&\rightarrow&\hat{\bar H}^{(j)}(t)=\hH^{(j})(t)+\sum_{l=0}^{M-1}
\sum_{\lp=0}^{M-1}\frac{g_{l\lp}(t)}{2\imath}\left(
\alpha^{(j)*}_{l\lp}(t)\hG^{(j)}_{l\lp}(t)-\alpha^{(j)}_{l\lp}(t)\hG^{(j)\dagger}_{l\lp}(t)\right)+\beta(t)\hid^{(j)},
\label{eq:transfLinbopb}
\eea 
\ese
\end{widetext}
with $\alpha^{(j)}_{l\lp}(t)$ and $\beta(t)$ arbitrary complex and real functions of time respectively.
This means that if we use any of the new set of Lindblad operators $\{\hE^{(j)}_{l\lp}(t)\}_{l,\lp=0,\ldots,N_j-1}$,
and the new Hamiltonians $\hat{\bar H}^{(j)}(t)$,  the dynamics of the reduced states $\hrho_j(t)$ are the same.
We stress that the transformations in Eqs.~\eqref{eq:transfLinbop} modify the unitary and non-unitary part of the evolution 
in the master equation Eq.~\eqref{eq:maserteqrho} in a nontrivial way.
Also, it is worth to mention that in order to obtain the invariance of the master equation Eq.~\eqref{eq:maserteqrho},  it is crucial to consider the null terms in Eq.~\eqref{eq:derirhojb}
(see Eq.~\eqref{eq:secondtermzero}).

\section{The canonical Hamiltonian}
\label{sec:ham}
%
The ambiguity in determining the unitary and non-unitary part of the evolution in a master equation in Lindblad form, due to transformations of the type in Eqs.~\eqref{eq:transfLinbop}, can be resolved within the framework developed in~\cite{HaydenSorce2022}, which is given deeper physical meaning in~\cite{colla2022open}.
\par
For our purposes two results of~\cite{HaydenSorce2022} are important. First, the authors show that, in finite-dimensional quantum systems, the reduced dynamics associated with a global evolution given by  a completely positive dynamical map is generically described by a master equation, \ie 
\bea
\label{eq:masterequation}
\dv{\hrho}{t}=\calL_t(\hrho(t)),
\eea
where the superoperator $\calL_t$ has the Lindblad form. Second, 
there is a unique Hamiltonian, called the canonical one, that describes the unitary part of evolution, associated with $\calL_t$, for which the non-unitary part of evolution, \ie the dissipative part, has a minimal size evaluated with a suitable norm. 
\par
These results were used in~\cite{colla2022open} to develop the principle of minimum dissipation that correspond to identify the canonical Hamiltonian as the unique local Hermitian operator that defines the internal energy of a subsystem whose evolution is described by a master equation. Interesting enough this canonical Hamiltonian is in general time dependent even if the global evolution is described by an autonomous Hamiltonian. So, there is a unique and consistent definition of work as we will see in Section~\ref{sec:thermo}.
\par 
In this section we will prove that the Hamiltonians $\hH^{(j)}(t)$, of each subsytem $j=1,2$, given in~\eqref{eq:defHjt}, are indeed the canonical ones. In doing so we start 
briefly summarizing the results in~\cite{HaydenSorce2022}.
\par
Let's first note that in any master equation, the generator $\calL_t$ is hermiticity-preserving and trace-annihilating,  \ie 
it satisfys that $\calL_t(\hM)^\dagger=\calL_t(\hM^\dagger)$ and 
$\Tr(\calL_t(\hrho(t)))=0$. This allows the decomposition
\beq
\label{eq:decommaseq}
\calL_t(\cdot)=\Phi_{\hH(t)}(\cdot)+\calD_t(\cdot),
\eeq
where $\Phi_{\hH(t)}$ is a Hamiltonian superopetor,
\beq
\label{eq:hamilsuperopr}
\Phi_{\hH(t)}(\cdot)=-\imath[\hH(t),\cdot],
\eeq
that describes the unitary part of the evolution, and $\calD_t$ is the dissipative superoperator that describes the non-unitary part of the evolution, which generically, in finite-dimension quantum systems, has the Lindblad form~\cite{HaydenSorce2022}. 
\par
All possible superoperators $\calL_t$ form a real vector space $\mathfrak{qme}(\calH)$ with a subspace 
$\mathfrak{ham}(\calH)$ of Hamiltonian  superoperators $\Phi_{\hH(t)}$ associated with Hermitian operators $\hH(t)$. In~\cite{HaydenSorce2022} is defined the inner product 
\bea
\label{eq:innerprodct}
\lala\calM|\calN\rara_{\text{avg}}= \overline{\matrixel{\psi}{\overline{\calM(\dyad{\phi})\calN(\dyad{\phi})}}{\psi}}, 
\eea
where overlines denote averages with respect to the Haar
measure $d\hU$ over the unitary group,
that allows an orthogonal decomposition $\mathfrak{qme}(\calH)=\mathfrak{ham}(\calH)\oplus \mathfrak{ham}(\calH)^{\perp}$. Therefore, 
for superoparators in~\eqref{eq:decommaseq}, we have $\Phi_{\hH(t)}\in \mathfrak{ham}(\calH)$ and 
$\calD_t\in \mathfrak{ham}(\calH)^{\perp}$. Within all possible orthogonal decompositions of $\calL_t$, there is an unique one such 
\bea
\label{eq:modDmin}
\|\calD_{t,\text{min}}\|_{avg}\leq \|\calD_t\|_{avg},
\eea
where $\|\cdot\|_{avg}$ is the norm defined from the innner product~\eqref{eq:innerprodct}. This means that $\calD_{t,\text{min}}$ has no component on $\mathfrak{ham}(\calH)$.
Furthermore, it has been proven that the canonical Hamiltonian $\hH_{\text{can}}$ associated with the 
Hamiltonian superoperator in the decomposition, 
$\calL_t(\cdot)=\Phi_{\text{can}}(\cdot)+\calD_{\text{min}}(\cdot)$, has always the form \cite{footnote1}:
\bea
\label{eq:Hcan2} 
\hH_{\text{can}}&=&\frac{1}{2\imath}\int d\hU\;(\hU^\dagger{\calL_t}(\hU)-{\calL_t}(\hU^\dagger)\hU).
\eea
\par
Now we define the linear map $\matPhi : \calL_t\rightarrow \matPhi(\calL_t)$ between superoperators
$\calL_t$ and operators $\matPhi(\calL_t)$:
\bea
\matPhi(\calL_t)=\frac{1}{2\imath}\int d\hU\;(\hU^\dagger{\calL_t}(\hU)-{\calL_t}(\hU^\dagger)\hU).
\eea
We denote $\Phi_{\hH^{(j)}(t)}$, $j=1,2$, the Hamiltonian superoperators in the master equations
in~\eqref{eq:maserteqrho}, with the Hamiltonians $\hH^{(j)}(t)$ defined in~\eqref{eq:defHjt},
and $\calD_t^{(j)}$ their dissipator superoparators defined in~\eqref{eq:derirhoj}. 
So, $\matPhi({\cal L}^{(j)})=\matPhi(\Phi_{\hH^{(j)}(t)})+\matPhi(\calD^{(j)})$. 
\par
In Appendix~\ref{AppA} we show that
for any Hamiltonian superoperator  
$\Phi_{\hH}$, with $\hH=\hH(t)$ acting on a Hilbert space ${\cal H}$ of dimension $d=\dim({\cal H})$, we have that 
\beq
\hH^{\prime}=\matPhi(\Phi_{\hH})=\hH-\frac{\Tr(\hH)}{d}\hid.
\label{eq:PhimapH}
\eeq
The dissipator superoperators $\calD_t^{(j)}$, in Eqs.~\eqref{eq:derirhoj}, verify that 
\beq
\matPhi(\calD_t^{(j)})=0.
\label{eq:PhimapDzero}
\eeq 
This follows from the fact that the Lindblad 
operators $\hG^{(j)}_{l\lp}(t)$, $j=1,2$,  have null trace.
\par
Therefore, we finally arrive to 
\bea
\hH^{(j)}_{\text{can}}(t)=\matPhi(\Phi_{\hH^{(j)}(t)})=\hH^{(j)}(t)-\frac{\Tr(\hH^{(j)}(t))}{N_j}\hid_j.
\label{eq:canHjt}
\eea
This shows  that the Hamiltonians $\hH^{(j)}(t)$, $j=1,2$, in the master equations in~\eqref{eq:maserteqrho} are the unique canonical Hamiltonians, 
except for the value of the trace that has no effect in the evolution of the reduced states $\hrho^{(j)}(t)$.
It is worth to note that 
\beq
\calD_{t,\text{min}}^{(j)}=\calD_t^{(j)},
\eeq
\ie the dissipator in~\eqref{eq:derirhoj} verifies the principle of minimum dissipation.
\par
The most general transformations that leave a Lindblad type master equation 
invariant \cite{footnote2}, is like the ones in Eqs.~\eqref{eq:transfLinbop}
but including a linear transformation of the sets of $(N^{(j)})^2$ operators $\{\sqrt{|g_{l\lp}(t)|}\hG_{l\lp}^{(j)}(t)\}$
, ordered as vectors, \ie
\beq
\label{eq:gaugeA}
\text{vec}\left(\{\hG_{l\lp}^{\prime(j)}(t)\}\right)=\AA(t)
\text{vec}\left(\{\sqrt{|g_{l\lp}(t)|}\hG_{l\lp}^{(j)}(t)\}\right),
\eeq
described by a time dependent $(N^{(j)})^2\times (N^{(j)})^2$ unitary matrix $\AA(t)$  that preserve a diagonal matrix $\JJ_t$
with $\text{sgn}(g_{l\lp}(t))$ at its entries, and $g_{l\lp}(t)$ is defined in Eq.~\eqref{eq:defgllpt}.
The set of transformations in Eqs.~\eqref{eq:transfLinbop}, together with~\eqref{eq:gaugeA}, obey a group property, so they constitute the symmetry group of the Lindblad master equation in consideration \cite{colla2022open}. Clearly, the transformation~\eqref{eq:gaugeA} preserve the null trace of the operators $\{\hG_{l\lp}^{\prime(j)}(t)\}$, so the use of the new set $\{\hG_{l\lp}^{(j)}(t)\}$ to describe 
the same dissipator superoperator $\calD_t^{(j)}$ does not change the result in Eq.~\eqref{eq:PhimapDzero}.
Thus, the Hamiltonians $\hH^{(j)}_{\text{can}}(t)$ in Eq.\eqref{eq:canHjt} remain invariant under the transformation~\eqref{eq:gaugeA}. 
\section{Open quantum system perspective}
\label{sec:open}

\par
So far, we have demonstrated that the master equation Eq. ~\eqref{eq:maserteqrho} (see also~\eqref{eq:derirhoj}) derived in Ref. \cite{malavazi2022schmidt}, using the Schmidt decomposition, adheres to the principle of minimum dissipation \cite{HaydenSorce2022}, with the unitary part corresponding to the canonical Hamiltonian. This raises the question of whether this master equation is unique or if alternative formulations exist that describe the evolution of the subsystems. In this section, we demonstrate from the perspective of open quantum systems that master equation Eq. ~\eqref{eq:maserteqrho} is indeed unique.
\par
Without lost of generality we focus on the dynamics of the reduced state of the  subsytem $1$, given by a completely positive dynamical map 
$\hrho^{(1)}(t)=\calN_t(\hrho_0^{(1)})$. We assume that the subsystem $2$ start in a pure state $\ket{\phi_0^{(2)}}$ and the global unitary evolution of both subsystems is given by $\hU(t)=e^{-\imath \hH t}$ with $\hH$ the Hamiltonian in~\eqref{eq:defH}. So, the superoperator $\calN_t$ is:
 \bse
 \label{eq:Krausrep0}
\bea
\calN_t(\ldots)&=&\Tr_2\left(\hU(t)\ket{\phi^{(2)}_0}\ldots\bra{\phi_0^{(2)}}\hU^\dagger(t)\right)
\label{eq:Krausrepa}\\
&=&\sum_{l=0}^{N_2-1} \hK_l(t) \ldots \hK^\dagger_l(t),
\label{eq:Krausrep}
\eea
\ese 
In the last equality we present the Choi-Kraus representations of the dynamic map with Kraus operators:
\bea
\hK_l(t)=\bra{\psi^{(2)}_l(t)}\hU(t)\ket{\phi_0^{(2)}}.
\eea
The different Kraus operators appear from the realization of the trace in ~\eqref{eq:Krausrepa} using arbitrary bases, $\{\ket{\psi^{(2)}_l(t)}\}_{l=0,\ldots,N_2-1}$, in subsystem $2$. It should be noted that since we allow arbitrary time dependence in the basis states $\ket{\psi^{(2)}_l(t)}$, this affects the time dependence of the Kraus operators $\hK_l(t)$, but not that of the superoperator $\calN_t$. Note also that $\calN_t$ can propagate any initial state $\hrho_0^{(1)}$, pure or mixed, of subsystem $1$, as long as subsystem $2$ always starts with the same pure state $\ket{\phi_0^{(2)}}$.
\par
Because we are dealing with finite dimensional Hilbert spaces, the linear map $\calN_t$ has possible null determinant only at isolated fixed points \cite{HaydenSorce2022}. Therefore, it must be invertible on an interval $[0,t^*)$, where $t^*$ is the first time that $\calN_t$ has zero determinant \cite{footnote3}. Thus, the master equation for the evolution of $\hrho^{(1)}(t)$
is,
\beq
\label{eq:mastereqLt}
\dv{\hrho^{(1)}(t)}{t}=\calL_t^{(1)}(\hrho^{(1)}(t))=\dot{\calN_t}(\calN_t^{-1}(\hrho^{(1)}(t))).
\eeq
In \cite{HaydenSorce2022} it was proved that master equations characterized by a superoperator
$\calL_t^{(1)}$, constructed in the manner shown in~\eqref{eq:mastereqLt} from a completely positive and time-continuous map
$\calN_t$, have Lindblad form. 
\par
The Lindblad form of the master equation~\eqref{eq:mastereqLt} is exact and not an approximation of any kind. It is also important to note that the superoperator $\calL_t^{(1)}$ depends on the initial condition $\hrho_0^{(1)}$ of the evolution, since $\dot{\calN_t}(\calN_t^{-1}(\hrho^{(1)}(t)))=\dot{\calN_t}(\hrho_0^{(1)})$. Therefore, for each initial condition we have a different master equation~\eqref{eq:mastereqLt}, \ie different superoperators $\calL^{(1)}$, although each initial condition can be propagated with the same dynamical map $\calN_t$. Interesting enough~\eqref{eq:mastereqLt} is the only exact master equation, associated with the dynamical map $\calN_t$, to propagate an initial state, pure or mixed, $\hrho_0^{(1)}$. Of course, because of its Linblad form, each master equation has its symmetry group.
\par 
Let's now find the master equation in ~\eqref{eq:mastereqLt}, with $\calN_t$ defined
in ~\eqref{eq:Krausrepa}, for a given initial pure state $\hrho_0^{(1)}=\dyad{\phi_0^{(1)}}$.
This can be done straightforward if we incorporate the trajectory information, in the space of density operators of subsystem $2$, of the reduced state $\hrho^{(2)}(t)$.
Indeed, once we specify an initial state of the entire bipartite system, the unitary evolution $\hU(t)$ uniquely determines the trajectories in the space of density operators of each subsystem, \ie $\hrho^{(j)}(t)$ with $j=1,2$. In the case where each subsystem starts in a pure state, the information of the trajectories $\hrho^{(j)}(t)$ is in their eigenstates that enter the Schmidt decomposition~\eqref{eq:SchDecOrg}. Therefore, we can define a particular faithful 
Choi-Kraus representations of $\calN_t$, with Kraus operators 
$\hKb_l(t)$ in~\eqref{eq:Krausrep} such $\ket{\psi_l^{(2)}(t)}=\ket{\phi_l^{(2)}(t)}$, with $l=0,\ldots,N_2-1$, being the eigenstates of $\hrho^{(2)}(t)$.
\par
Note that the Kraus operators $\hKb_l(t)$ verifies:
\beq
\hKb_l(t)\hrho_0^{(1)}=\left\{
   \begin{matrix} 
      s_l(t)\hU^{(1)}(t)\hG^{(1)}_{l0}(0)& \text{if $l\leq M-1$} \\
      0 & \text{if $l> M-1$} \\
   \end{matrix}\right.,
   \label{eq:Krausprime}
\eeq
when $\hrho_0^{(1)}$ is pure, which follows immediately from the Schmidt decomposition in ~\eqref{eq:SchDecBritosimpl}, and from the fact that $\ket{\phi_l^{(2)}(t)}$, for $l>M-1$ are eigenstates with zero eigenvalue.
\par
Thus, using the property~\eqref{eq:Krausprime} and the fact that $\hrho_0^{(1)2}=\hrho_0^{(1)}$, we can write 
\begin{widetext}
\bea
\hrho^{(1)}(t)=\calN_t(\hrho_0^{(1)})=\sum_{l=0}^{M-1} \hKb_l(t) \hrho_0^{(1)} \hKb^{\dagger}_l(t)
=\sum_{l=0}^{M-1} s_l^2(t) \hU^{(1)}(t) \hG^{(1)}_{l0}(0) \hrho_0^{(1)} \hG^{(1)\dagger}_{l0}(0)
\hU^{(1)\dagger}(t)
=
\sum_{l=0}^{M-1} s_l^2(t) \hG^{(1)}_{ll}(t),
\label{eq:Schimdt2}
\eea   
\end{widetext}
that it is the spectral decomposition in~\eqref{eq:hrhojta} that comes from the Schmidt decomposition of $\hU(t)\ket{\phi_0^{(1)}}\ket{\phi_0^{(2)}}$ in~\eqref{eq:SchDecBritosimpl}.
It is important to realize that, contrary to the set $\{\hKb_l(t) \}_{l=0,\ldots,N_2-1}$, the set of operators $\{s_l(t) \hU^{(1)}(t) \hG^{(1)}_{l0}(0)\}_{l=0,\ldots,M-1}$ in~\eqref{eq:Schimdt2} does not constitute a faithful Choi-Kraus representation of $\calN_t$ in~\eqref{eq:Krausrepa}. Therefore, it is only the property~\eqref{eq:Krausprime} of the superoperators $\hK_l(t)$, that allows us to calculate from~\eqref{eq:Schimdt2} the inverse map:
\begin{widetext}
\bea
\hrho_0^{(1)}=\calN_t^{-1}(\hrho^{(1)}(t))=\sum_{l=0}^{M-1}\frac{1}{M s^2_l(t)}\hG_{l0}^{(1)\dagger}(0)\hU^{(1)\dagger}(t)
\hrho^{(1)}(t) \hU^{(1)}(t)\hG_{l0}^{(1)}(0).
\eea
\end{widetext}
Note that, because $s_l(t)>0$,  $\calN_t^{-1}$ is well define for all $t\geq0$, becoming the identity map when $t=0$.
\par
We can also compute $\dv{\hKb_l(t)}{t}\hrho_0^{(1)}$ from~\eqref{eq:Krausprime} to evaluate $\dv{\hrho^{(1)}(t)}{t}=\dot{\calN}_t(\calN_t^{-1}(\hrho^{(1)}(t)))$ and then recover the master equation in~\eqref{eq:maserteqrho} in a Lindblad form (see Eqs.~\eqref{eq:derirhoj}) that corresponds to a minimal dissipator part.
Therefore, we have proved that this master equation is the only exact master equation that describes the evolution of reduced states, starting from pure product states, in bipartite interacting autonomous systems in general.

%
\section{Heat and work in the subsytems }
\label{sec:thermo}
%
In sections~\eqref{sec:ham} and~\eqref{sec:open}, we showed that the Hamiltonians $\hH^{(j)}(t)$
in~\eqref{eq:defHjt}, that appear in the master equations in~\eqref{eq:maserteqrho}, are indeed the canonical Hamiltonians so they can be used to uniquely identify  what it is work and heat in each of the subsystems  \cite{colla2022open}. 
Also, can be used to formulate a first law of quantum thermodynamics.
We will expand on this below.
\par
The conservation of energy in the bipartite autonomous system considered is (see Eq. ~\eqref{eq:relbetdiagel} in Appendix~\ref{AppB}):
\beq
\expval{\hH^{(1)}(t)}_{\hrho^{(1)}(t)}+\expval{\hH^{(2)}(t)}_{\hrho^{(2)}(t)}=\expval{\hH}_{0},
\label{eq:totalenergy}
\eeq
with $\expval{\hH}_{0}=\Tr(\hrho_0^{(1)}\otimes \hrho_0^{(2)}\;\hH)$ ($\hrho_0^{(j)}=\dyad{\phi_0^{(j)}}$, $j=1,2$, are the initial states) being
the initial energy disposal in the whole system, and 
$\expval{\hH^{(j)}(t)}_{\hrho^{(j)}(t)}=\Tr(\hrho^{(j)}(t)\hH^{(j)}(t))$, $j=1,2$,
are the internal energies of the susbsytems. Throughout the evolution the balance in the flux of the whole energy is given by   
\bea
\label{eq:fluxbalance}
\dv{\expval{\hH^{(1)}(t)}_{\hrho^{(1)}(t)}}{t}=-\dv{\expval{\hH^{(2)}(t)}_{\hrho^{(2)}(t)}}{t}
\eea
where the rate of change of the internal energies is  
 \bea
 \label{eq:firstthermlaw}
\dv{\expval{\hH^{(j)}(t)}_{\hrho^{(j)}(t)}}{t}=\dv{Q^{(j)}(t)}{t}+\dv{W^{(j)}(t)}{t}.
\eea 
Following the useful definitions of the rate of change of heat and work in each subsystem as~\cite{alicki2018introduction},
\begin{widetext}
\bse
\bea
\dv{Q^{(j)}(t)}{t}&=&\Tr\left(\dv{\hrho^{(j)}(t)}{t}\hH^{(j)}(t)\right)
=\sum_{l=0}^{M-1}\dv{s^2_l(t)}{t}\matrixel{\phi_l^{(j)}(t)}{\hH^{(j)}(t)}{\phi_l^{(j)}(t)}
\label{eq:Qjtexpsum}\\
\dv{W^{(j)}(t)}{t}&=&\Tr\left(\hrho^{(j)}(t)\dv{\hH^{(j)}(t)}{t}\right)
=
\sum_{l=0}^{M-1}s^2_l(t)
\left(\matrixel{\phi_l^{(j)}(t)}{
\dv{\hH^{(j)}(t)}{t}}{\phi_l^{(j)}(t)}\right)\nonumber\\
&=&
\sum_{l=0}^{M-1}s^2_l(t)\dv{}{t}\left(\matrixel{\phi_l^{(j)}(t)}{\hH^{(j)}(t)}{\phi_l^{(j)}(t)}\right),
\label{eq:Wjtexpsum}
\eea
\ese
\end{widetext}
respectively. Expression~\eqref{eq:Qjtexpsum}  is obtained using~\eqref{eq:maserteqrho} and Eq.~\eqref{eq:relbetdiagel}
in Appendix \ref{AppB}, while expression~\eqref{eq:Wjtexpsum} is obtained using~\eqref{eq:hrhojt}, Eq.~\eqref{eq:relbetdiagel}
in Appendix \ref{AppB} and that 
$\matrixel{\phi_l^{(j)}(t)}{\dv{\hH^{(j)}(t)}{t}}{\phi_l^{(j)}(t)}=\dv{}{t}\left(\matrixel{\phi_l^{(j)}(t)}{\hH^{(j)}(t)}{\phi_l^{(j)}(t)}\right)$.
\par
Using that $\dv{s^2_l(t)}{t}=2s_l(t)\dv{s_l(t)}{t}$ with
\bea
\dv{s_l(t)}{t}
&=&\Im\left(\bra{\phi^{(1)}_l(t)}\bra{\phi^{(2)}_l(t)}\lambda\hV\ket{\Psi(t)}\right)
\nonumber\\
&=&\Im(z_l(t)),
\label{eq:dvdts2}
\eea
proved in Appendix~\ref{AppB}, we recognize from~\eqref{eq:Qjtexpsum} that the rate of heat change is directly proportional to the strength of the interaction controlled by $\lambda$, being zero for $\lambda=0$. Note, also that the local effective Hamiltonians $\hH^{(j)}(t)$, $j=1,2$, are related to the free autonomous Hamiltonians $\hH^{(j)}$ in~\eqref{eq:defH} through, $\hH^{(j)}(t)=\hH^{(j)}+\hH^{\prime(j)}(t)$~\cite{malavazi2022schmidt}, with
$\hH^{(j)}(t)=\hH^{(j)}$ in the absence of interaction. So in this case
we also have that the rate of change of work is zero. When the interaction is turned on
the contribution of the free autonomous Hamiltonians $\hH^{(j)}$ to the rate of change of work is only through their contribution to the evolution of the reduced states
$\hrho^{(j)}(t)$.
\par
Note that Eqs.~\eqref{eq:firstthermlaw} correspond to the statement of the first law of thermodynamics in each of the subsystems $j=1,2$.
This allows to rewrite~\eqref{eq:fluxbalance} as 
\bea
\dv{\left(\sum_{j=1}^2\,Q^{(j)}(t)\right)}{t}=-\dv{\left(\sum_{j=1}^2\,W^{(j)}(t)\right)}{t}.
\eea
In turn, using Eq.~\ref{eq:relbetdiagel} in Appendix \ref{AppB}, that the rate of change of the total amount of heat in the bipartite system can be written as 
 \bea
\dv{\left(\sum_{j=1}^2Q^{(j)}(t)\right)}{t}=\sum_{j=1}^2\left(\dv{Q_F^{(j)}(t)}{t}\right)
+\dv{Q_{\text{int}}(t)}{t},
\eea
where 
\beq
\dv{Q_F^{(j)}(t)}{t}=\sum_{l=0}^{M-1}\dv{s^2_l(t)}{t}\matrixel{\phi_l^{(j)}(t)}{\hH^{(j)}}{\phi_l^{(j)}(t)},
\eeq 
j=1,2, are the rate of change of the heat associated 
 with the free Hamiltonians $\hH^{(j)}$, $j=1,2$, in Eq.~\eqref{eq:defH}. 
The contribution from the interaction,
\bea
\dv{Q_{\text{int}}(t)}{t}
&=&\Im\left(\sum_{l=0}^{M-1} z^2_l(t)\right).
\eea
can be obtained using also the equality $2\Im(z_l(t))\Re(z_l(t))=\Im(z^2_l(t))$, with $z_l(t)$ defined in~\eqref{eq:dvdts2}.
For the total amount of work in the bipartite system, using Eq.~\ref{eq:relbetdiagel} in Appendix \ref{AppB}, we arrive to  
 \bea
\dv{\left(\sum_{j=1}^2W^{(j)}(t)\right)}{t}&=&\dv{W_{\text{int}}(t)}{t}\nonumber\\
&=&\sum_{l=0}^{M-1}s_l(t)
\Re\left(\dv{z_l(t)}{t}\right)\nonumber\\
&&-\frac{1}{2}\Im\left(\sum_{l=0}^{M-1} z^2_l(t)\right).
\eea

\section{Conclusions}
\label{sec:conclusions}

\par

One of the most promising general frameworks for quantum thermodynamics of open systems is based on the exact, time-convolutionless quantum master equation, which describes the interaction between a system and its environment without approximations, and incorporates the principle of minimum dissipation. This principle asserts that, since master equations typically take the Lindblad form, there is a unique decomposition of the system's evolution into reversible (unitary) and non-reversible (dissipative) components, where the dissipator is minimized. This allows us to identify the genuine local internal energy as the one associated with the so-called canonical Hamiltonian, which governs the unitary part of the minimal dissipation decomposition. This Hamiltonian is generally time-dependent and influenced by the interaction energy between the system and the bath.
Consequently, the principle of minimum dissipation provides a physically justified method for calculating thermodynamic quantities such as work and heat.

In this work, we advance the effort to establish a general quantum thermodynamic theory by uncovering the connection between the Schmidt decomposition of the evolution of an initial product of pure states and the principle of minimum dissipation in general interacting bipartite autonomous systems. Specifically, we demonstrate that the master equations governing the evolution of reduced states, derived from the Schmidt decomposition, is the one dictated by the principle of minimum dissipation. As a result, the time-dependent canonical Hamiltonian is the one that generates the local unitary transformation between the local Schmidt bases at any given time and the initial basis.

Moreover, through the use of the completely positive dynamical map formalism and its Kraus-Choi representations, we show that the unique master equations found within this formalism are indeed those arising from the Schmidt decomposition. Lastly, we analyze the definitions of heat and work that emerge from this formalism showing that the rate of heat change is directly proportional to the
interaction strength.
\acknowledgments{ FT acknowledges financial support from the Brazilian agency INCT-Informa\c{c}\~ao Qu\^antica and CAPES-Print. 
This work has been partially supported by  CONICET (Grant No.~PIP
11220200100568CO), UBACyT (Grant  No.~ 20020170100234BA and 20020220300049BA) and ANPCyT
(PICT-2020-SERIEA-01082). We thanks Augusto Roncaglia and Esteban A. Calzetta for valuable comments on the manuscript.
}
\appendix
\section{Proof of Eq.~(\ref{eq:PhimapH})}
\label{AppA}
Eq.~\eqref{eq:PhimapH} can be proven writing, 
\bea
\matPhi(\Phi_{\hH})&=&-\frac{1}{2}\int d\hU\;(\hU^\dagger[\hH,\hU]-[\hH,\hU^\dagger]\hU)\nonumber\\
&=&-\frac{1}{2}\int d\hU\;(\hU^\dagger (\hH\hU-\hU\hH)-(\hH\hU^\dagger-\hU^\dagger\hH)\hU)\nonumber\\
&=&-\int d\hU\; (\hU^\dagger\hH\hU-\hH)=\hH-\int d\hU\; \hU^\dagger\hH\hU \nonumber\\
&=&\hH-\frac{\Tr(\hH)}{d}\hid,
\eea
where $d=\dim(\calH)$, and  we use the identity, 
\beq
\int d\hU\; \hU^\dagger\hM\hU=\frac{\Tr(\hM)}{d}\hid,
\eeq 
for any operator $\hM$ acting on $\calH$.
%
\section{A useful identities}
\label{AppB}
In this appendix, we prove equalities used in Sec. \ref{sec:thermo}. Using the Schmidt decomposition in Eq.~\eqref{eq:SchDecOrg}, we have that, 
\bea
\bra{\phi^{(2)}_l(t)}\ket{\Psi(t)}=s_l(t)\ket{\phi^{(1)}_l(t)}.
\eea
Taking its time derivative and making the overlap with $\bra{\phi_{\lp}^{(1)}(t)}$, on both sides,
we arrive to,
\begin{widetext}
\bea
-\bra{\phi^{(1)}_{\lp}(t)}\bra{\phi^{(2)}_l}\hH_2(t)\ket{\Psi(t)}+\bra{\phi^{(1)}_{\lp}(t)}\bra{\phi^{(2)}_l}\hH\ket{\Psi(t)}&=&
\imath \dv{s_l(t)}{t} \delta_{l\lp}+s_l(t)\bra{\phi^{(1)}_{\lp}(t)}\hH_1(t)\ket{\phi^{(1)}_l(t)}
\label{eq:intermidiateeq}
\eea
\end{widetext}
where $\hH$ is the Hamiltonian in Eq. \eqref{eq:defH}.
Now, using again the Schmidt decomposition for $\ket{\Psi(t)}$, we evaluate the matrix elements,
\begin{widetext}
\bea
\bra{\phi^{(1)}_{\lp}(t)}\bra{\phi^{(2)}_l(t)}(\hH_2-\hH_2(t))\ket{\Psi(t)}&=&\sum_k s_k(t) \bra{\phi^{(2)}_l(t)}(\hH_2-\hH_2(t))\ket{\phi_k^{(2)}(t)}
\bra{\phi^{(1)}_{\lp}(t)}\ket{\phi_k^{(1)}(t)}\nonumber\\
&=&s_{\lp}(t)\matrixel{\phi^{(2)}_l(t)}{(\hH_2-\hH_2(t))}{\phi^{(2)}_{\lp}(t)}\\
\bra{\phi^{(1)}_{\lp}(t)}\bra{\phi^{(2)}_l(t)}\hH_1\ket{\Psi(t)}&=&
s_l(t)\matrixel{\phi^{(1)}_{\lp}(t)}{\hH_1}{\phi^{(1)}_l(t)}
\eea
\end{widetext}
Replacing these matrix elements in Eq. \eqref{eq:intermidiateeq} and rearranging, finally we arrive to,
\begin{widetext}
\bea
&&s_l(t)\bra{\phi^{(1)}_{\lp}(t)}(\hH_1(t)-\hH_1)\ket{\phi^{(1)}_l(t)}
+s_{\lp}(t)\matrixel{\phi^{(2)}_l(t)}{(\hH_2(t)-\hH_2)}{\phi^{(2)}_{\lp}(t)}\nonumber\\
&=&\bra{\phi^{(1)}_{\lp}(t)}\bra{\phi^{(2)}_l(t)}\lambda\hV\ket{\Psi(t)}-\imath \dv{s_l(t)}{t} \delta_{l\lp}.
\label{eq:identitymain1}
\eea
\end{widetext}
When $l=\lp$, because the l.h.s. of Eq.~\eqref{eq:identitymain1} is real, if we conjugate all the equation we arrive to~\eqref{eq:dvdts2}.
Using this result in Eq. \eqref{eq:identitymain1}, for  $l=\lp$ we get 
\begin{widetext}
\bea
\bra{\phi^{(1)}_l(t)}\hH_1(t)\ket{\phi^{(1)}_l(t)}
+\matrixel{\phi^{(2)}_l(t)}{\hH_2(t)}{\phi^{(2)}_l(t)}
&=&\bra{\phi^{(1)}_l(t)}\hH_1\ket{\phi^{(1)}_l(t)}
+\matrixel{\phi^{(2)}_l(t)}{\hH_2}{\phi^{(2)}_l(t)}
\nonumber\\
&&+
\frac{\Re(z_l(t))}{s_l(t)},
\label{eq:relbetdiagel}
\eea
\end{widetext}
with $z_l(t)$ defined in~\eqref{eq:dvdts2}.
From Eq.~\eqref{eq:relbetdiagel} and 
Eq.~\eqref{eq:maserteqrho} it  immediately follows 
Eq.~\eqref{eq:totalenergy}.
\bibliographystyle{apsrev4-2}
\bibliography{main3}

\begin{thebibliography}{17}
\bibitem [{foo({\natexlab{a}})}]{footnote0}%
  \BibitemOpen
  \href@noop {} {\bibinfo {title} {Note that the preparation of pure initial states are generally the goal in controlled experimental situations consisting of interacting systems of few degrees of freedom.}} ({\natexlab{a}})\BibitemShut {NoStop}%
\bibitem [{foo({\natexlab{b}})}]{footnote1}%
  \BibitemOpen
  \href@noop {} {\bibinfo {title} {In the demonstration it is important to realize that the inner product $ \lala\phi_{\hat{h}_l}|\phi_{\hH_{\lp}} \rara_{\text{avg}}$ between any two elements of $\mathfrak{ham}(\cal{H})$ is proportional to the hilbert-schmidt inner product, \ie $\lala\phi_{\hH_l}|\phi_{\hH_{\lp}}\rara_{\text{avg}}=\frac{2}{d(d+1)}\tr( {\hH}_l {\hH}_{\lp})$ where $d=\dim(\cal{H})$.}} ({\natexlab{b}})\BibitemShut {NoStop}%
\bibitem [{foo({\natexlab{c}})}]{footnote2}%
  \BibitemOpen
  \href@noop {} {\bibinfo {title} {In our case the master in eq.~\eqref{eq:maserteqrho}}} ({\natexlab{c}})\BibitemShut {NoStop}%
\bibitem [{foo({\natexlab{d}})}]{footnote3}%
  \BibitemOpen
  \href@noop {} {\bibinfo {title} {$t^*$ is infinite when ${\calN}_t$ is invertible everywhere.}} ({\natexlab{d}})\BibitemShut {NoStop}%
\end{thebibliography}


\end{document}